\documentclass[11pt, leqno]{article}
\usepackage{amsmath}
\usepackage{amsthm}
\usepackage{amssymb}
\usepackage{latexsym}

\allowdisplaybreaks[1]
\usepackage{amsfonts}
\usepackage{ascmac}
\usepackage{color}
\usepackage{enumerate}

\usepackage{graphicx}

\usepackage[T1]{fontenc}
\theoremstyle{definition}
\theoremstyle{plain}

\allowdisplaybreaks[1]

\date{}
\newcommand{\dis}{\displaystyle}

\newcommand{\D}{\mbox \DH}

\def\text#1{\mbox{#1 }}

\title{\bf Thermodynamically consistent modeling for dissolution/growth of bubbles in an incompressible solvent}
\author{Dieter Bothe
\footnote{ Technische Universit\"at Darmstadt, Center of Smart Interfaces,  
64287 Darmstadt,  Germany\qquad (bothe@csi.tu-darmstadt.de).} 
\,\, and \,\, Kohei Soga
\footnote{CNRS-ENS Lyon, UMPA UMR 5669, 69364 Lyon cedex 7, France    
 (kohei.soga@ens-lyon.fr).}}
\begin{document}
\maketitle
\begin{abstract}
\noindent  We derive mathematical models of the elementary process of dissolution/growth of bubbles in a liquid under pressure control. The modeling starts with a fully compressible version, both for the liquid and the gas phase so that the entropy principle can be easily evaluated. This yields a full PDE system for a compressible two-phase fluid with mass transfer of the gaseous species. Then the passage to an incompressible solvent in the liquid phase is discussed,  where a carefully chosen equation of state for the liquid mixture pressure allows for a limit in which the solvent density is constant. We finally provide a simplification of the PDE  system in case of a dilute solution.     
\medskip

\noindent{\bf Keywords:} \medskip two-phase fluid system; mass transfer; entropy principle; incompressible limit

\end{abstract}
\setcounter{section}{0}
\setcounter{equation}{0}
\section{Introduction}

The process of dissolution or growth of gas bubbles in an ambient liquid phase is very common in many situations. In everyday life, we often see  bubbles in carbonated mineral water, beer, champagne etc.
In particular the dissolution of gases is of huge technological and industrial importance in the context of gas scrubbing. 
This is, for instance, relevant for CO$_2$ disposal, where gas from a combustion process is injected into a reactive liquid medium. 
Such processes are usually run under pressure control instead of volume control. Note that the latter is much more common in the mathematical analysis of such mass transfer problems, since it allows for a fixed domain in which the mathematical model--usually in the form of a system of partial differential equations--holds. The massive impact of the external pressure is known from the above mentioned everyday life examples, but also can be seen in the medical context. This is the case with decompression sickness or caisson disease, where severe symptoms can be caused by bubble generation in the blood after a fast change of the ambient pressure. 
There is a large literature on experiments and numerical computation of dissolution/growth of bubbles in a liquid, e.g. Liger-Belair et al. \cite{Liger-Belair}, Sauzade and Cubaud \cite{Sauzade and Cubaud}, Takemura and Yabe \cite{Takemura and Yabe}.  A rigorous mathematical model is necessary for possible theoretical investigations and mathematical analysis on this topic.  

Based on Continuum Physics, we derive a mathematical model of a two-phase fluid system of type liquid/gas, where both gas and liquid phases are composed of molecularly miscible constituents and the pressure is controlled via a free (upper) surface $\Gamma(t)$. The system consists of chemical components $A_1,\ldots,A_N$.  The gas phase is denoted by $\Omega^+(t)$, the liquid phase by $\Omega^-(t)$ and the movable free interface by $\Sigma(t)$. See the Figure 1 below.    
\begin{center}
\includegraphics[trim = 0 260 0 240, width=10cm]{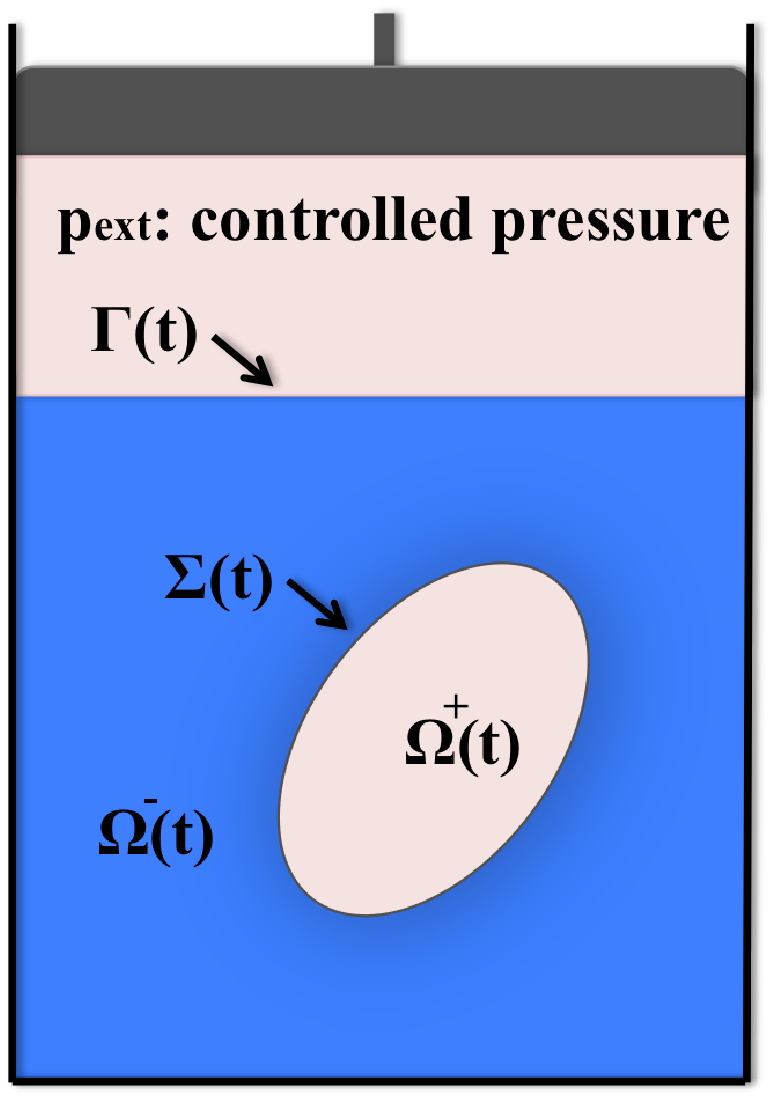}\\
Figure 1. The two phase system under pressure control.
\end{center}
\indent In common mathematical models for mass transfer from or to gas bubbles in a liquid phase, the transferred gas is treated as a dilute component in both phases. This allows to use a two-phase Navier-Stokes system together with advection-diffusion equations for passive scalars. If the bubble is composed of a pure gas, this is no longer possible since the dissolution then significantly changes the bubble volume. In this case a much more elaborate modeling is required for both of the bulk phases and the transmission condition at the interface. In particular, the two one-sided limits of the bulk velocities at the interface and the interface's own velocity need to be distinguished. Since such a more rigorous model accounts for the mass and volume of individual constituents, an incompressible model for the liquid phase will still lead to non-zero divergence of the barycentric velocity field. Moreover, a thermodynamically rigorous model needs to be developed for compressible bulk phases in the first place. Only then, an incompressible version may be derived as a limit, where the latter depends on the notion of incompressibility which is neither a priori clear nor unique in the mixture context. 

\indent The novel aspect in the present paper is the idea of an incompressible solvent (associated to $A_N$) carrying dissolved gas components which add their partial pressure to the total one  like being ideal gases. The underlying mixture is supposed to be described by an equation of state according to 
\begin{eqnarray*}                   
p=p^R_N+K(\frac{\rho_N}{\rho_N^R}-1)+\sum_{k=1}^{N-1}\frac{\rho_k}{M_k}RT,
\end{eqnarray*}
where $p^R_N$ is a reference pressure and $\rho_N^R$ a reference density for the solvent, while $K$ is the solvent bulk modulus. The incompressible limit will be attained (formally) by letting $K$ tend to infinity. This leads to the constraint 
\begin{eqnarray*}
\rho_N\equiv \rho_N^R,
\end{eqnarray*}    
i.e.\ to a constant solvent density. Since the continuity equation for the solvent then reduces to 
\begin{eqnarray*}   
\nabla\cdot v_N=0,
\end{eqnarray*}
it makes sense to employ the solvent momentum balance instead of the one for the mixture. This is attractive, because it leads to a standard incompressible Navier-Stokes equation for the bulk liquid.  Only the diffusive fluxes, which rely on the relative velocity to the barycentric one, become slightly more intricate, but only involving a simple linear relation. 

The obtained PDE systems still comprise of a compressible gas phase model. Low Mach number approximation seems possible and will be given in a forthcoming  paper. Note that the gas phase density in the incompressible limit will still be a function of time, determined by the dynamical mass transfer process.   
\setcounter{section}{1}
\setcounter{equation}{0}
\section{Balance Equations}
\noindent{\bf (i) Mass balance} 

For simplicity, we  assume that there are no chemical reactions (which could be easily added) and that there is no absorbed mass at the interface, i.e.\ $\rho_i^\Sigma\equiv 0$ for all $i=1,\ldots,N$. The partial mass balance in its integral form for a fixed control volume $V$ with the outer normal $n$ reads as  
\begin{eqnarray*}
&& \frac{d}{dt}\int_V \rho_i\, dx=-\int_{\partial V} \rho_i v_i \cdot n\,do.
\end{eqnarray*}
Using the two-phase transport and divergence theorems (see the appendix), this implies 
\begin{eqnarray*}
\int_{V\setminus \Sigma}\partial _t\rho_i\,dx-\int_{\Sigma_V}[\![\rho_i]\!] v^\Sigma\cdot n_\Sigma \,do=-\int_{V\setminus\Sigma}\nabla\cdot(\rho_iv_i)\,dx-\int_{\Sigma_V}[\![\rho_iv_i\cdot n_\Sigma]\!]\,do
\end{eqnarray*}
with $\Sigma_V:=\Sigma(t)\cap V$, the surface velocity $v^\Sigma$ and the surface unit normal $n_\Sigma$ pointing toward $\Omega^-$. 
Comparison of bulk and interface terms yields the local form
\begin{eqnarray}\label{(1)}
\left\{
\begin{array}{lll}
&\dis \partial_t\rho_i+\nabla\cdot(\rho_iv_i)=0\,\,\,\,\mbox{in $\Omega^+(t)\cup\Omega^-(t)$,}\medskip\\
&[\![\rho_i(v_i-v^\Sigma)\cdot n_\Sigma]\!]=0\,\,\,\,\mbox{on $\Sigma(t)$}.
\end{array}
\right.
\end{eqnarray}
Above, the bracket $[\![\,\,]\!]$ denotes the jump of a quantity across the interface (crossing $\Sigma$ in the direction opposite to $n_\Sigma$).
The mixture is described by the total density $\rho$ and the barycentric velocity $v$, given by 
\begin{eqnarray*}
\rho:=\sum_{i=1}^N\rho_i,\quad \rho v:=\sum_{i=1}^N\rho_iv_i.
\end{eqnarray*}
As a consequence of (\ref{(1)}), the mixture obeys the continuity equation 
\begin{eqnarray}\label{(2)}
\left\{
\begin{array}{lll}
&\dis \partial_t\rho+\nabla\cdot(\rho v)=0\,\,\,\,\mbox{in $\Omega^+(t)\cup\Omega^-(t)$,}\medskip\\
&[\![\rho(v-v^\Sigma)\cdot n_\Sigma]\!]=0\,\,\,\,\mbox{on $\Sigma(t)$}.
\end{array}
\right.
\end{eqnarray}
Let $\dot{m}^\pm$ denote the one-sided limits $\rho^\pm(v^\pm-v^\Sigma)\cdot n_\Sigma$ on $\Sigma(t)$.  Then the second equation in (\ref{(2)}) becomes $\dot{m}^-=\dot{m}^+$, and hence $\dot{m}:=\dot{m}^-=\dot{m}^+$ is well-defined.  Similarly, we introduce $\dot{m}_i:=\dot{m}_i^\pm=\rho_i^\pm(v_i^\pm-v^\Sigma)\cdot n_\Sigma$. We define diffusion velocities $u_i:=v_i-v$, mass fractions $y_i:=\rho_i/\rho$ and diffusion mass fluxes $j_i:=\rho_iu_i=\rho_i(v_i-v)$. Then we have the following equivalent form of the equations (\ref{(1)}):    
\begin{eqnarray}\label{(1)'}
\left\{
\begin{array}{lll}
&\dis \partial_t\rho_i+\nabla\cdot(\rho_iv+j_i)=0\,\,\,\,\mbox{in $\Omega^+(t)\cup\Omega^-(t)$,}\medskip\\
&[\![j_i\cdot n_\Sigma]\!]+[\![\rho_i(v-v^\Sigma)\cdot n_\Sigma]\!]=0\,\,\,\,\mbox{at $\Sigma(t)$},
\end{array}
\right.
\end{eqnarray}
or
\begin{eqnarray}\label{(1)''}
\left\{
\begin{array}{lll}
&\dis \rho(\partial_ty_i+v\cdot\nabla y_i)+\nabla \cdot j_i=0\,\,\,\,\mbox{in $\Omega^+(t)\cup\Omega^-(t)$,}\medskip\\
&[\![j_i\cdot n_\Sigma]\!]+\dot{m}[\![y_i]\!]=0\,\,\,\,\mbox{at $\Sigma(t)$}.
\end{array}
\right.
\end{eqnarray}
In the common models for mass transfer, the jump condition in (\ref{(1)'}) or (\ref{(1)''}) is simplified to read $[\![ j_i\cdot n_\Sigma]\!]=0$, assuming $\dot{m}=0$ which means that the total phase change effect of the mass transfer is neglected; cf. Bothe and Fleckenstein \cite{Bothe and Fleckenstein}  for an assessment of this approximation. 

\noindent {\bf (ii) Momentum balance}

 The mixture is to be described by a so-called class-I model, where we consider only a single (common) momentum balance. The integral form is 
 \begin{eqnarray*}
 \frac{d}{dt}\int_V\rho v \,dx&=&-\int_{\partial V}\rho v(v\cdot n)\,do+\int_{\partial V}Sn\,do+\int_V\rho b\,dx+\int_{\partial\Sigma_V}S^\Sigma \nu\,ds 
 \end{eqnarray*}
 with the bulk stress tensor $S$, the surface stress tensor $S^\Sigma$ and the body force $\rho b$. Note that $\rho b=\sum_{k=1}^N\rho_kb_k$ with (possibly) individual body forces $b_k$, for instance due to forces in an electrical field.   Here $\nu$ is the outer unit normal of the bounding curve $\partial \Sigma_V$ of $\Sigma_V$, being tangential to $\Sigma$. The transport and (surface) divergence theorems yield the local form  
\begin{eqnarray}\label{(3)} 
\left\{
\begin{array}{lll}
&\dis \partial_t(\rho v)+\nabla \cdot (\rho v\otimes v- S)=\rho b\,\,\,\,\mbox{in $\Omega^+(t)\cup\Omega^-(t)$,}\medskip\\
&\dot{m}[\![v]\!]-[\![Sn_\Sigma]\!]=\nabla_\Sigma \cdot S^\Sigma\,\,\,\,\mbox{at $\Sigma(t)$}.
\end{array}
\right.
\end{eqnarray}
We assume non-polar fluids, for which the balance of angular momentum has a simple form without body couples or surface couples. This is equivalent to the assumptions
\begin{eqnarray*}
S=S^{\sf T}, \quad S^\Sigma=(S^\Sigma)^{\sf T}. 
\end{eqnarray*}  
This is a constitutive assumption which is made right away. 

\noindent {\bf (iii) Energy balance}

The integral form of the total energy balance is 
\begin{eqnarray*}
&&\frac{d}{dt}\left[ \int_V \rho(e+\frac{v^2}{2})\,dx+\int_{\Sigma_V}u^\Sigma\,do \right]
=-\int_{\partial V}\rho (e+\frac{v^2}{2})v\cdot n\, do-\int _{\partial \Sigma_V}u^\Sigma v^\Sigma\cdot \nu \,ds\\
&&\qquad-\int_{\partial V}q\cdot n \,do -\int_{\partial \Sigma_V}q^\Sigma\cdot \nu \,ds + \int_{\partial V} v\cdot Sn\,do+\int_{\partial \Sigma_V}v^\Sigma\cdot S^\Sigma \nu\, ds \\
&&\qquad+\int_{V} v\cdot \rho b\,dx+\int_V \sum_{k=1}^N j_k\cdot b_k\,dx
\end{eqnarray*} 
with the specific internal energy of the bulk $e$ and the internal energy density of the surface $u^\Sigma$. After straightforward computations, the local form turns out as
\begin{eqnarray*}\label{(4)} 
\,\,\,\,\left\{
\begin{array}{lll}
&\dis \partial_t\left(\rho(e+\frac{v^2}{2})\right)+\nabla \cdot \left(\rho (e+\frac{v^2}{2})v+q\right)=\medskip \\ 
&\dis\qquad\qquad\qquad\qquad\qquad\qquad\qquad  \nabla\cdot (Sv)+\rho v\cdot b+\sum_{k=1}^N j_k\cdot b_k \quad\mbox{ in $\Omega^+(t) \cup\Omega^-(t)$,}\medskip\\
&\dis \frac{D^\Sigma u^\Sigma}{Dt}+u^\Sigma\nabla_\Sigma \cdot v^\Sigma +[\![\rho(e+\frac{v^2}{2})(v-v^\Sigma)\cdot n_\Sigma]\!]+[\![q\cdot n_\Sigma]\!]+\nabla_\Sigma\cdot q^\Sigma=\medskip \\ 
&\qquad\qquad \qquad\qquad\qquad\qquad\qquad\qquad [\![Sv\cdot n_\Sigma]\!]+\nabla_\Sigma \cdot (S^\Sigma v^\Sigma )\,\,\,\,\mbox{  at $\Sigma(t)$}.
\end{array}
\right.
\end{eqnarray*}
Subtracting the balance of kinetic energy derived from (\ref{(3)}), one obtains the balance of internal energy as 
\begin{eqnarray}\label{(5)} 
\,\,\,\,\left\{
\begin{array}{lll}
&\dis \partial_t(\rho e)+\nabla \cdot (\rho e v+q)=\nabla v:S +\sum_{k=1}^N j_k\cdot b_k
\quad\mbox{in $\Omega^+(t)\cup\Omega^-(t)$,}\medskip\\
&\dis \frac{D^\Sigma u^\Sigma}{Dt}+u^\Sigma\nabla_\Sigma\cdot v^\Sigma +\dot{m}[\![e+\frac{(v-v^\Sigma)^2}{2}-\frac{1}{\rho}n_\Sigma\cdot Sn_\Sigma]\!]+[\![q\cdot n_\Sigma]\!]\\
& \qquad\qquad\qquad\qquad+\nabla_\Sigma \cdot q^\Sigma- [\![(v-v^\Sigma)_\parallel\cdot Sn_\Sigma]\!]=\nabla_\Sigma v^\Sigma :S^\Sigma\,\,\,\,\mbox{at $\Sigma(t)$},
\end{array}
\right.
\end{eqnarray}
where $(v-v^\Sigma)_\parallel$ stands for the tangential projection of $(v-v^\Sigma)$ onto the local tangent plane to $\Sigma$, i.e.\ $(v-v^\Sigma)_\parallel=P_\Sigma(v-v^\Sigma)$ with the projection tensor $P_\Sigma=I-n_\Sigma\otimes n_\Sigma$.  
Later, we will use the constitutive relation $S^\Sigma=\gamma^\Sigma P_\Sigma$ with a scalar $\gamma^\Sigma$. Then we have $\nabla_\Sigma v^\Sigma:S^\Sigma=\gamma^\Sigma\nabla_\Sigma \cdot v^\Sigma$. 

\noindent{\bf (iv) Entropy balance} 

Let $\rho s$ denote the density of entropy in the bulk (i.e.\ $s$ is the specific entropy) and $\eta^\Sigma$ the area-density of interfacial entropy. 
The integral form of the entropy balance is  
\begin{eqnarray*}
\frac{d}{dt} \left[ \int_V\rho s\,dx + \int_{\Sigma_V}\eta^\Sigma\,do \right]&=&
-\int_{\partial V}(\rho s v +\Phi)\cdot n\,do - \int_{\partial \Sigma_V} (\eta^\Sigma v^\Sigma+\Phi^\Sigma)\cdot \nu\,ds \\
&& + \int_{V}\xi\,dx +\int_{\Sigma_V} \xi^\Sigma\,do
\end{eqnarray*}
with the bulk entropy flux $\Phi$ and the interfacial entropy flux $\Phi^\Sigma$. 
Hence we obtain the local form 
\begin{eqnarray}\label{(6)} 
\,\,\,\,\left\{
\begin{array}{lll}
&\dis \partial_t(\rho s)+\nabla \cdot (\rho s v+\Phi)=\xi 
\quad\mbox{in $\Omega^+(t)\cup\Omega^-(t)$,}\medskip\\
&\dis \frac{D^\Sigma \eta^\Sigma}{Dt}+\eta^\Sigma\nabla_\Sigma \cdot v^\Sigma +\dot{m}[\![s]\!]+\nabla_\Sigma\cdot \Phi^\Sigma+[\![\Phi\cdot n_\Sigma]\!]=\xi^\Sigma\,\,\,\,\mbox{at $\Sigma(t)$}.
\end{array}
\right.
\end{eqnarray}
\vspace{-5mm}
\setcounter{section}{2}
\setcounter{equation}{0}
\section{Entropy Principle}
If the entropy fluxes $\Phi$ and $\Phi^\Sigma$ in (\ref{(6)}) are related to the primitive variables via constitutive relations in such a way that the following entropy principle holds, we speak of a thermodynamically consistent model.  
\medskip
\medskip

\noindent{\bf Entropy principle.} {\it The entropy flux ($\Phi,\Phi^\Sigma$) is such that 
\begin{itemize}
\item The entropy production is a sum of binary products of ``fluxes'' times ``driving force'', i.e. $\dis \xi=\sum_m F_m D_m$ and $\dis \xi^\Sigma=\sum_{m'} F^\Sigma_{m'} D_{m'}^\Sigma$. 
\item $\xi\ge0$, $\xi^\Sigma\ge 0$ for any thermodynamical process.
\item $\xi\equiv 0$ and $\xi^\Sigma\equiv 0$ characterizes equilibria of the system.   
\end{itemize}}
\medskip
\medskip

\noindent This is a condensed form of the full entropy principle. For more details see Bothe and Dreyer \cite{Bothe and Dreyer}, as well as Dreyer \cite{Dreyer}. We consider the simplest class of isotropic fluids without mesoscopic forces. This corresponds to the choice of certain primitive variables in modeling the entropy of the material. We assume  
\begin{eqnarray}\label{(7)}
\rho s=h(\rho e,\rho_1,\ldots,\rho_N),\quad\eta^\Sigma=h^\Sigma(u^\Sigma),
\end{eqnarray}
where $h$ and $h^\Sigma$ are concave functions. The concavity  is required for thermodynamic stability properties of the mixture. Then we define the (absolute) temperature $T$, respectively $T^\Sigma$ of bulk and interface, as well as the bulk chemical potentials $\mu_i$ via 
\begin{eqnarray}\label{(8)} 
\frac{1}{T}:=\frac{\partial h}{\partial (\rho e)},\,\,\,\,\,-\frac{\mu_i}{T}:=\frac{\partial h}{\partial \rho_i},\,\,\,\,\,\frac{1}{T^\Sigma}:=\frac{\partial h^\Sigma}{\partial u^\Sigma}. 
\end{eqnarray}
Next, we compute $\xi$ and $\xi^\Sigma$ from (\ref{(6)}), (\ref{(7)}), (\ref{(8)}), where we eliminate the derivatives of $\rho_i$, $\rho e$, $u^\Sigma$ by means of the balance equations in (\ref{(1)}), (\ref{(5)}). This yields the following results. 

\noindent {\bf (i) Bulk entropy production}
\begin{eqnarray*}
\xi&=& \nabla\cdot(\Phi - \frac{q}{T}+\sum_{k=1}^N\frac{\mu_kj_k}{T})-\frac{1}{T}(\rho e -\rho s T-\sum_{k=1}^N \rho_k \mu_k)\nabla \cdot v\\
&&+\frac{1}{T}\nabla v:S+q\cdot \nabla\frac{1}{T}-\sum_{k=1}^Nj_k\cdot\left(\nabla\frac{\mu_k}{T}-\frac{b_k}{T}\right).
\end{eqnarray*} 
We choose the entropy flux as  
\begin{eqnarray*}
\Phi=\frac{q}{T}-\sum_{k=1}^N\frac{\mu_kj_k}{T}
\end{eqnarray*}
and determine further constitutive relations so that the entropy principle holds. We decompose the stress tensor $S$ as $S=-PI+S^\circ$ with the traceless part $S^\circ$ of $S$ and $P=-\frac{1}{3}\mbox{tr}(S)$. We decompose the pressure $P$ as $P=p+\Pi$, where $\Pi$ vanishes in equilibrium. This is important, since $\Pi$ can depend on $\nabla\cdot v$, while $p$ cannot. Hence $S$ is rewritten as 
\begin{eqnarray*}
S=-(p+\Pi)I+S^\circ.
\end{eqnarray*}
Introducing the Helmholtz free energy 
\begin{eqnarray*}
\rho\psi=\rho\psi(T,\rho_1,\ldots,\rho_N)=\rho e-\rho sT,
\end{eqnarray*}
we change from $\rho e$ as a primitive variable to $T$ (via Legendre transform with $\frac{\partial (\rho e)}{\partial (\rho s)}=T$). Then $\xi$ becomes
\begin{eqnarray*}
\xi =-\frac{1}{T}\left( \rho \psi + p-\sum_{k=1}^N\rho_k\mu_k \right)\nabla\cdot v - \frac{\Pi}{T}\nabla\cdot v+\frac{1}{T}\nabla v :S^\circ + q \cdot \nabla \frac{1}{T}-\sum_{k=1}^Nj_k\cdot \left(\nabla\frac{\mu_k}{T}-\frac{b_k}{T}\right).
\end{eqnarray*}
Now, $\xi\ge0$ for any thermodynamical process implies the Gibbs-Duhem relation 
\begin{eqnarray*}\label{(9)}
p&=&-\rho\psi+\sum_{i=k}^N\rho_k\mu_k.
\end{eqnarray*}
Thus the entropy production in the bulk reduces to read
\begin{eqnarray*}
\label{(10)}
\xi&=&-\frac{\Pi}{T}\nabla \cdot v +\frac{1}{T}\nabla v: S^\circ+q\cdot \nabla \frac{1}{T} 
-\sum_{k=1}^Nj_k\cdot \left(\nabla\frac{\mu_k}{T}-\frac{b_k}{T}\right). 
\end{eqnarray*}
Since $\xi\ge0$ is required, the simplest closure is linear in the driving forces and such that a quadratic form is obtained. Note that the constraint $\sum_{k=1}^N j_k=0$ has to be accounted for. Hence we eliminate $j_N$, which is chosen as $-\sum_{k=1}^{N-1} j_k$. For $D:=\frac{1}{2}(\nabla v+(\nabla v)^{\sf T})$ and its traceless part $D^\circ$ we have $\nabla v:S^\circ=D^\circ:S^\circ$ and $\mbox{tr}D=\nabla \cdot v$. Then $\xi$ becomes 
\begin{eqnarray}\label{xi}   
\xi=-\frac{\Pi}{T} \nabla\cdot v +\frac{1}{T}D^\circ:S^\circ+q\cdot\nabla \frac{1}{T}-\sum_{k=1}^{N-1} j_k\cdot\left(\nabla\frac{\mu_k-\mu_N}{T}-\frac{b_k-b_N}{T}\right).
\end{eqnarray}
Note that the viscous entropy production can be written as $\frac{1}{T}D:S^{irr}$, if we let $S^{irr}:=-\Pi I+S^\circ$, i.e.\ $S^{irr}$ is the irreversible part of $S$ which produces entropy.

\noindent{\bf (ii) Interfacial entropy production}

\noindent  We do not consider viscous surface dissipation, hence $S^\Sigma=\gamma^\Sigma P_\Sigma$. Then it follows from the second equation in (\ref{(6)}) and the other balance equations that 
\begin{eqnarray*}
\xi^\Sigma&=&\frac{1}{T^\Sigma}(\gamma^\Sigma-u^\Sigma+T^\Sigma\eta^\Sigma)\nabla_\Sigma\cdot v^\Sigma+\nabla_\Sigma\cdot(\Phi^\Sigma-\frac{q^\Sigma}{T^\Sigma})+q^\Sigma\cdot\nabla_\Sigma\frac{1}{T^\Sigma}\\
&&+[\![ (\frac{1}{T}-\frac{1}{T^\Sigma})(\dot{m}sT+q\cdot n_\Sigma) ]\!]+\frac{1}{T^\Sigma}[\![ (v-v^\Sigma)_\parallel\cdot(S^{irr}n_\Sigma) ]\!]
- \sum_{k=1}^N[\![ \frac{\mu_kj_k\cdot n_\Sigma}{T} ]\!]\\
&&-\frac{\dot{m}}{T^\Sigma}[\![ \sum_{k=1}^Ny_k\mu_k+\frac{(v-v^\Sigma)^2}{2}-\frac{1}{\rho}n_\Sigma\cdot S^{irr}n_\Sigma ]\!].
\end{eqnarray*}
We choose the entropy flux as 
\begin{eqnarray*}
\Phi^\Sigma=q^\Sigma/T^\Sigma
\end{eqnarray*}
and obtain the surface Gibbs-Duhem equation
\begin{eqnarray*}
\gamma^\Sigma=u^\Sigma-T^\Sigma\eta^\Sigma,
\end{eqnarray*}
which shows that $\gamma^\Sigma$ is the interfacial free energy. For simplification, we assume from here on that there is no temperature jump at $\Sigma(t)$, i.e. 
\begin{eqnarray*}
[\![T]\!]=0,\quad T|_{\Sigma}=T^\Sigma. 
\end{eqnarray*}
Then, with (\ref{(1)'}), we see that $\xi^\Sigma$ becomes 
\begin{eqnarray}\label{(11)} 
\xi^\Sigma&=&q^\Sigma\cdot\nabla_\Sigma \frac{1}{T}+\frac{1}{T}[\![ (v-v^\Sigma)_\parallel\cdot(S^{irr} n_\Sigma) ]\!]\\\nonumber
&&\qquad\qquad\qquad\qquad-\frac{1}{T}\sum_{k=1}^N\dot{m}_k[\![ \mu_k+\frac{(v-v^\Sigma)^2}{2}-\frac{1}{\rho}n_\Sigma\cdot S^{irr}n_\Sigma]\!],
\end{eqnarray}
where $\dot{m}_i$ satisfies $[\![\dot{m}_i]\!]=0$ for all $i=1,\ldots,N$. 

In the next section, we further determine appropriate constitutive relations such that the entropy principle holds. In addition, one needs a constitutive modeling for the Helmholtz free energy $\rho\psi$. This will be constructed from an equation of state for the pressure $p$ and from  the chemical potentials $\mu_i$. 
\setcounter{section}{3}
\setcounter{equation}{0}
\section{Constitutive Modeling}
Constitutive relations can be derived from the entropy principle in (\ref{xi}) and (\ref{(11)}). The standard closure is as follows (cf.\ de Groot and Mazur \cite{de Groot and Mazur}; Slattery \cite{Slattery}; Hutter and J\"ohnk \cite{Hutter and Johnk}).  
\medskip
\medskip

\noindent {\bf (i) Bulk}
\begin{itemize}
\item[(B1)] $\Pi=-\lambda\nabla\cdot v, \,\,\,\lambda=\lambda(T,\rho_i)\ge 0$ the bulk viscosity,
\item[(B2)] $S^\circ=2\eta D^\circ,\,\,\,\eta=\eta(T,\rho_i)\ge 0$ the dynamic viscosity (Newton's law), 
\item[(B3)] $q=\alpha \nabla\frac{1}{T},\,\,\,\alpha=\alpha(T,\rho_i)\ge0$ the heat conductivity (Fourier's law),  
\item[(B4)] $j_i=-\sum_{k=1}^{N-1} L_{ik}\left(\nabla\frac{\mu_k-\mu_N}{T}-\frac{b_k-b_N}{T}\right)$ with a positive (semi-)definite matrix \\$[L_{ik}]=[L_{ik}(T,\rho_1,\ldots,\rho_N)]$ of mobilities (Fick's law for multi-component mixture). 
\end{itemize}

\noindent {\bf (ii) Interface}
\begin{itemize}
\item[(B5)] $q^\Sigma=\alpha^\Sigma\nabla_\Sigma\frac{1}{T}$, $\alpha^\Sigma=\alpha^\Sigma(T)\ge 0$ the interfacial heat conductivity,
\item[(B6)] $[\![ v_\parallel ]\!]=0$, $v_\parallel^\pm=v_\parallel^\Sigma$, i.e.\  continuous tangential velocities, 
\item[(B7)] If $i\in I^\pm:=\{i\,|\,A_i \mbox{ is only in }\Omega^\pm \}$, then $\dot{m}_i=0$ (no transfer) and otherwise 
$$ [\![ \mu_i]\!]=[\![ \frac{1}{\rho}n_\Sigma\cdot S^{irr}n_\Sigma-\frac{(v-v^\Sigma)^2}{2}]\!],$$  
or, more general but still neglecting mass transfer cross-effects, 
\item[(B7')]  $\dis \dot{m}_i=-\beta_i[\![  \mu_i+\frac{(v-v^\Sigma)^2}{2}-\frac{1}{\rho}n_\Sigma\cdot S^{irr}n_\Sigma ]\!]$, $\beta_i=\beta_i(T)\ge0$.
\end{itemize}

\indent Now we model the Helmholtz free energy $\rho\psi$, where we follow Example 2 in Bothe and Dreyer \cite{Bothe and Dreyer}.  The free energy can be constructed from an equation of state for the pressure $p$  and from relations for the ``chemical part'' of the chemical potential $\mu_i$. We consider the gas phase as an ideal mixture of ideal gases and the liquid phase as a solution with $A_N$ as the solvent and $A_1,\ldots,A_{N-1}$ the solutes (i.e. dissolved components). We introduce the following notation:
\begin{eqnarray*}  
c_i:=\frac{\rho_i}{M_i}\mbox{ (molar density)},\,\,\,c:=\sum_{i=1}^Nc_i,\,\,\,x_i:=\frac{c_i}{c}\mbox{ (molar fraction)},\,\,\,x':=(x_1,\ldots,x_{N-1}),
\end{eqnarray*}
where $\sum_{k=1}^{N}x_k=1$. We use $(\rho,x')$ as a set of primitive variables as well as $(\rho_1,\ldots,\rho_N)$. Note that $(\rho,x')\mapsto (\rho_1,\ldots,\rho_N)$ is one-to-one with the relations above and
\begin{eqnarray*}
\rho_i=\rho_i(\rho,x'):=\frac{\rho M_ix_i}{M(x')},\quad M(x'):=\sum_{k=1}^NM_kx_k,\quad x_N:=1-\sum_{k=1}^{N-1}x_k.
\end{eqnarray*}
Each thermodynamic quantity $f$ is represented as 
\begin{eqnarray*} 
f=f(T,\rho_1,\ldots,\rho_N)=\tilde{f}(T,\rho,x'),
\end{eqnarray*}
where we always suppose the above relations among $(T,\rho,x')$, $x_N$ and $(T,\rho_1,\ldots,\rho_N)$. 

Now we model the pressure. In the gas phase $\Omega^+(t)$, we assume $p=\sum_{k=1}^Np_k$ with partial pressures $p_i$ according to the ideal gas law $p_i=\frac{\rho_i}{M_i}RT$, namely  
\begin{eqnarray}\label{p_g}
p=p(T,\rho_1,\ldots,\rho_N)=\sum_{k=1}^N\frac{\rho_i}{M_i}RT=\tilde{p}(T,\rho,x')=\frac{\rho R T}{M(x')},
\end{eqnarray} 
where $\rho_i=0$ means that $A_i$ does not exist in $\Omega^+(t)$. 

In the liquid phase $\Omega^-(t)$, for the later passage to the incompressible case, we use 
\begin{eqnarray}\label{revise1}
p_N=p_N^R+K(\frac{\rho_N}{\rho_N^R}-1) 
\end{eqnarray}
with a bulk modulus $K=\partial_{\rho_N}p_N(\rho_N^R)\rho_N^R>0$ and reference quantities $p_N^R$ and $\rho_N^R$. Later we let $K\to \infty$, which leads to $\rho_N\equiv\rho_N^R$. Note that the ``$1$'' in (\ref{revise1}) can be generalized to an appropriate function of temperature and composition, but then $\rho_N$ will not become constant in the incompressible limit. For all other species in the liquid, we assume that they behave as ideal gas components (in the solvent ``matrix'' instead of a gas volume), namely $p_i=\frac{\rho_i}{M_i} RT$ for all $i<N$. Hence we have 
\begin{eqnarray}\label{p_l}
p&=&p(T,\rho_1,\ldots,\rho_N)=p_N^R+K(\frac{\rho_N}{\rho_N^R}-1) + \sum_{k=1}^{N-1}\frac{\rho_k}{M_k}RT\\\nonumber
&=& \tilde{p}(T,\rho,x')= p_N^R+K\left(\frac{\rho M_N x_N}{\rho_N^R M(x')}-1\right) + \frac{\rho R T}{M(x')}\sum_{k=1}^{N-1}x_k,
\end{eqnarray}
where $\rho_i=0$ ($i<N$) means that $A_i$ does not exist in $\Omega^-(t)$. 

\indent The full chemical potential cannot be modeled directly, but needs to be computed from a Helmholtz free energy function $\rho\psi$. The modeling of $\psi$ follows the concept laid out in Section 13 of Bothe and Dreyer \cite{Bothe and Dreyer} and employs a decomposition of $\psi$ into an ``elastic'' part $\psi^{el}$, which takes into account the mechanical (pressure) work, and a ``thermal'' part $\psi^{th}$ which accounts for the entropy of mixing. 

We start with a fixed temperature $T$ and a reference pressure $p^R$.  We have a  reference density function $\rho^\ast=\rho^\ast(T,x')$ through the equation
\begin{eqnarray*} 
\tilde{p}(T,\rho^\ast,x')=p^R.
\end{eqnarray*}
We then define 
\begin{eqnarray*}
\psi^{th}(T,x'):=\tilde{\psi}(T,\rho^\ast(T,x'),x'), \,\,\,\,\, \psi^{el}(T,\rho,x'):=\tilde{\psi}(T,\rho,x')-\psi^{th}(T,x'). 
\end{eqnarray*}
Note that $\psi^{el}(T,\rho^\ast,x')=0$. From the Gibbs-Duhem relation, we obtain 
\begin{eqnarray*}
\rho^\ast\psi^{th}(T,x')=p^R+\sum_{k=1}^N\rho_k(\rho^\ast,x')\mu_k^{th}(T,x'),\,\,\,\mu_k^{th}(T,x'):=\tilde{\mu}_k(T,\rho^\ast(T,x'),x').
\end{eqnarray*}
The thermal part of the chemical potential needs to be modeled, where we only consider the case of  ideal mixtures (only containing entropy of mixing), namely
\begin{eqnarray*}
\mu^{th}_i(T,x')=g_i(T,p^R)+\frac{RT}{M_i} \ln x_i,\,\,\,i=1,\ldots,N,
\end{eqnarray*}
where $g_i$ denotes the Gibbs free energy of the pure component $A_i$ in the respective phase.  Next we compute $\psi^{el}$ through the relation 
\begin{eqnarray*}
 \frac{\partial}{\partial \rho}\psi^{el}(T,\rho,x')=\frac{\tilde{p}(T,\rho,x')}{\rho^2},  
\end{eqnarray*} 
inserting $\tilde{p}$ modeled in (\ref{p_g}) and (\ref{p_l}), respectively. 

For the gas phase,  we obtain 
\begin{eqnarray*}  
\psi^{el}(T,\rho,x')=\int^\rho_{\rho^\ast}\frac{\tilde{p}(T,\tilde{\rho},x')}{\tilde{\rho}^2}\,d\tilde{\rho}= \frac{RT}{M(x')}\ln \frac{\rho}{\rho^\ast}.
\end{eqnarray*}
Hence we have 
\begin{eqnarray*}
\rho\psi=\rho\tilde{\psi}(T,\rho,x')=-p^R\frac{\rho}{\rho^\ast}+\frac{\rho}{\rho^\ast}\sum_{k=1}^N\rho_k(\rho^\ast,x')\left\{g_k(T,p^R)+\frac{RT}{M_k}\ln x_k\right\}+\frac{\rho RT}{M(x')}\ln \frac{\rho}{\rho^\ast}.
\end{eqnarray*}
In oder to compute $\rho\psi(T,\rho_1\ldots,\rho_N)=\rho\tilde{\psi}(T,\rho,x'(\rho_1,\ldots,\rho_N))$, we observe the following relations: 
\begin{eqnarray}\nonumber  
&&\frac{\tilde{p}(T,\rho,x')}{\tilde{p}(T,\rho^\ast,x')}=\frac{\rho^\ast RT/M(x')}{\rho RT/M(x')}=\frac{\rho^\ast}{\rho}=\frac{p}{p^R}=\frac{RT}{p^R}\sum_{k=1}^N\frac{\rho_k}{M_k},\\\label{M}
&& M(x'(\rho_1\ldots,\rho_N))=\frac{\rho}{c},\,\,\,\,\,x_i=\frac{\rho_i/M_i}{ \sum_{k=1}^N\rho_k/M_k},\\\label{M2}
&&\rho_i(\rho^\ast(T,x'(\tilde{\rho}_1\ldots,\tilde{\rho}_N)),x'(\tilde{\rho}_1\ldots,\tilde{\rho}_N))=\rho^\ast
\frac{M_ix_i}{M(x')}=\rho^\ast\frac{\tilde{\rho}_i}{\tilde{\rho}}.
\end{eqnarray}
Direct calculation yields 
\begin{eqnarray*}
\rho\psi &=&\rho\psi(T,\rho_1\ldots,\rho_N)=\rho\tilde{\psi}(T,\rho,x'(\rho_1,\ldots,\rho_N))\\
&=&-RT\sum_{k=1}^N\frac{\rho_k}{M_k}+ \sum_{k=1}^N\rho_k\left(g_k(T,p^R)+\frac{RT}{M_k}\ln\frac{\rho_k}{M_k}\right) + RT\left(\sum_{k=1}^N\frac{\rho_k}{M_k}\right)\ln\frac{RT}{p^R}.
\end{eqnarray*}
Hence we obtain for $i=1,\ldots, N$ the chemical potentials as 
\begin{eqnarray*}
\mu_i=\mu_i(T,\rho_1,\ldots,\rho_N):=\frac{\partial (\rho\psi(T,\rho_1,\ldots,\rho_N))}{\partial \rho_i}
=g_i(T,p^R)+\frac{RT}{M_i}\ln\left( \frac{\rho_iRT}{p^RM_i} \right).
\end{eqnarray*}
 With the relation $\rho_iRT/M_i=(RT/M_i)(\rho M_ix_i/M(x'))=\tilde{p}(x,\rho,x')x_i$, we also obtain 
\begin{eqnarray}\label{(4.5)}
\mu_i=\tilde{\mu}_i(T,\rho,x') 
=g_i(T,p^R)+\frac{RT}{M_i}\ln \frac{\tilde{p}(T,\rho,x')}{p^R}+\frac{RT}{M_i}\ln x_i\mbox{ \,\,\,\,for $i=1,\ldots,N$}.
\end{eqnarray}
This reproduces the formulas known from the thermodynamical literature; see, e.g., M\"uller \cite{Muller}.  

For the liquid phase, we obtain 
\begin{eqnarray*}  
\psi^{el}(T,\rho,x')&=&\int^\rho_{\rho^\ast}\frac{\tilde{p}(T,\tilde{\rho},x')}{\tilde{\rho}^2}d\tilde{\rho}\\
&=& -(p^R_N-K)\left( \frac{1}{\rho}-\frac{1}{\rho^\ast}\right)+\frac{KM_Nx_N}{\rho^R_NM(x')}\ln\frac{\rho}{\rho^\ast}+\frac{RT}{M(x')}\left(\sum_{k=1}^{N-1}x_k\right)\ln\frac{\rho}{\rho^\ast}. 
\end{eqnarray*}
Hence we have 
\begin{eqnarray*}
\rho\psi&=&\rho\tilde{\psi}(T,\rho,x')=-p^R\frac{\rho}{\rho^\ast}+\frac{\rho}{\rho^\ast}\sum_{k=1}^N\rho_k(\rho^\ast,x')\left(g_k(T,p^R)+\frac{RT}{M_k}\ln x_k\right)\\
&&+(p^R_N-K)\left( \frac{\rho}{\rho^\ast}-1\right)+\frac{K\rho M_Nx_N}{\rho^R_NM(x')}\ln\frac{\rho}{\rho^\ast}+\frac{\rho RT}{M(x')}\left(\sum_{k=1}^{N-1}x_k\right)\ln\frac{\rho}{\rho^\ast}.
\end{eqnarray*}
Solving $\tilde{p}(T,\rho^\ast,x')=p^R$ with (\ref{M}), we get 
\begin{eqnarray*} 
\frac{\rho}{\rho^\ast}=\frac{\rho}{\rho^\ast(T,x'(\rho_1,\ldots,\rho_N))}=\left(K\frac{\rho_N}{\rho_N^R}+RT\sum_{k=1}^{N-1}\frac{\rho_i}{M_i} \right)(p^R-p^R_N+K)^{-1}.
\end{eqnarray*}
Straightforward computation with (\ref{M}) and (\ref{M2}) yields 
\begin{eqnarray*}
\rho\psi &=&\rho\psi(T,\rho_1\ldots,\rho_N)=\rho\tilde{\psi}(T,\rho,x'(\rho_1,\ldots,\rho_N))\\
&=&\left( RT\sum_{k=1}^{N-1}\frac{\rho_k}{M_k}+K\frac{\rho_N}{\rho_N^R} \right)\left( \ln \frac{\rho}{\rho^\ast} -1\right)+ K-p^R_N+\sum_{k=1}^N\rho_k\left(g_k(T,p^R)+\frac{RT}{M_k}\ln\frac{\rho_k/M_k}{c}\right),
\end{eqnarray*}
where the above $\rho/\rho^\ast$ and $c$ still have to be plugged in. Therefore we obtain, for $i=1,\ldots, N-1$,
\begin{eqnarray}\label{(4.6)}
\mu_i&=&\mu_i(T,\rho_1,\ldots,\rho_N):=\frac{\partial (\rho\psi(T,\rho_1,\ldots,\rho_N))}{\partial \rho_i}\\\nonumber
&=&g_i(T,p^R)+\frac{RT}{M_i}\ln\frac{\rho_i/M_i}{c} + \frac{RT}{M_i}\ln \left(\Big(K\frac{\rho_N}{\rho_N^R}+RT\sum_{k=1}^{N-1}\frac{\rho_k}{M_k} \Big)\frac{1}{p^R-p^R_N+K}\right).
\end{eqnarray}
For $i=N$, we obtain 
\begin{eqnarray}\label{(4.7)}
\quad\mu_N&=&\mu_N(T,\rho_1,\ldots,\rho_N):=\frac{\partial (\rho\psi(T,\rho_1,\ldots,\rho_N))}{\partial \rho_N}\\\nonumber
&=&g_N(T,p^R)+\frac{RT}{M_N}\ln\frac{\rho_N/M_N}{c} + \frac{K}{\rho^R_N}\ln \left(\Big(K\frac{\rho_N}{\rho_N^R}+RT\sum_{k=1}^{N-1}\frac{\rho_i}{M_i} \Big)\frac{1}{p^R-p^R_N+K}\right).
\end{eqnarray}
Let us sum up: Up to here, the balance equations (\ref{(1)}), (\ref{(3)}) and (\ref{(4)}) with constitutive relations (B1) to (B7), where the chemical potentials are modeled via (\ref{(4.5)}), (\ref{(4.6)}) and (\ref{(4.7)}), form -- up to  boundary and initial conditions -- a thermodynamically consistent full PDE system for a two-phase gas/liquid multicomponent system with  compressible liquid and gas phases and mass transfer.  For the non-isothermal case, the temperature dependencies need to be specified and the internal energy balance is usually transformed into a temperature form, i.e.\ of heat equation type. In the isothermal case, it can be dropped. 
\setcounter{section}{4}
\setcounter{equation}{0}
\section{Incompressible Limit}

We discuss the passage to an incompressible limit for the liquid solvent. As $K\to \infty$, assuming that the pressure stays bounded, we get $\rho_N/\rho^R_N\to1$. After a (formal) computation, the passage $K\to\infty$ yields $\mu_i\to\mu_i^\infty$, where 
 \begin{eqnarray*}\label{revise2}
\mu_i^\infty&=&g_i(T,p^R)+\frac{RT}{M_i}\ln x_i\mbox{\,\,\,\,\, for $i<N$},\\\label{revise3}
\mu_N^\infty&=&g_N(T,p^R)+\frac{p-p^R}{\rho^R_N}+\frac{RT}{M_N}\ln x_N.
\end{eqnarray*}
Note that for an incompressible pure substance $A_N$, the Gibbs free energy satisfies 
\begin{eqnarray*}
g_N(T,p)=g_N(T,p^R)+\frac{p-p^R}{\rho_N}.
\end{eqnarray*}
Hence we have
\begin{eqnarray*}
\mu_N^\infty=g_N(T,p)+\frac{RT}{M_N}\ln x_N.
\end{eqnarray*}
Therefore, we obtain the usual formulas for the chemical potential in the limit of $K\to\infty$, except for the fact that the chemical potentials of the solutes do not depend on the pressure. This is not a priori clear. Below, the superscript ``$\infty$'' is dropped. 

Note that $\rho_N$ is constant and $p=p_N+\sum_{k=1}^{N-1}\frac{\rho_k}{M_k}RT$ with $p_N$ a free primitive variable. In fact, $p_N$ acts as a Lagrange multiplier in the liquid phase to account for the constraint $\nabla \cdot v_N=0$ which results from (\ref{(1)}) for $i=N$.  As mentioned in the introduction, we employ the solvent momentum balance in the liquid phase and couple it to the barycentric momentum balance in the gas phase. For this purpose we use the relation 
\begin{eqnarray}\label{v_N-v}  
v_N=v+u_N=v+\frac{j_N}{\rho_N}=v-\frac{1}{\rho_N}\sum_{k=1}^{N-1}j_k.
\end{eqnarray}
Then each mass balance equation in (\ref{(1)}) is rewritten with $v_N$, instead of $v$,  in the liquid phase. In particular, the mass transfer transmission conditions $[\![ \dot{m}_i]\!]=0$ become, for $i<N$, 
\begin{eqnarray*}
\left(  j^+_i+\rho^+_i(v^+-v^\Sigma) \right)\cdot n_\Sigma=\left( J^-_i+\rho^-_i(v^-_N-v^\Sigma) \right)\cdot n_\Sigma\mbox{\quad on $\Sigma(t)$} ,
\end{eqnarray*}
where
\begin{eqnarray*}
J_i:=\rho_i(v_i-v_N)=j_i+\frac{\rho_i}{\rho_N}\sum_{k=1}^{N-1}j_k
\end{eqnarray*}
is the diffusion flux relative to the solvent. For $i=N$, the transfer condition is rewritten to become a substitution for the second equation in (\ref{(2)}) and reads as 
\begin{eqnarray}\label{jump-v_N}
j^+_N\cdot n_\Sigma + \dot{m}y^+_N \cdot n_\Sigma =\rho^-_N(v^-_N-v^\Sigma)\cdot n_\Sigma \mbox{\quad on $\Sigma(t)$}.
\end{eqnarray}
If the solvent evaporation is neglected, i.e.\ $\dot{m}_N=0$ and $\rho^+_N=0$, then (\ref{jump-v_N}) simplifies to 
\begin{eqnarray*}\label{jump-v_n2}
v^\Sigma\cdot n_\Sigma =v^-_N\cdot n_\Sigma\mbox{\quad on $\Sigma(t)$}.
\end{eqnarray*}
As for the momentum balance, the standard approach would be to employ the barycentric momentum balance (\ref{(3)}). However, this would lead to a velocity field $v$ of non-zero divergence. As an interesting alternative which leads to a divergence free velocity field, we make use of the partial momentum balance for $A_N$. According to Bothe and Dreyer \cite{Bothe and Dreyer}, the partial momentum balance for $A_i$ reads as 
\begin{eqnarray}\label{sharppp}
\rho_i(\partial_tv_i+v_i\cdot\nabla v_i)=-\rho_i\nabla\mu_i+\nabla\cdot S_i^{irr}+\rho_ib_i-T\sum_{k=1}^{N}f_{ik}\rho_i\rho_k(v_i-v_k),
\end{eqnarray}
where $S_i^{irr}=p_iI+S_i$ is the irreversible part of $S_i$ and $f_{ik}=f_{ki}>0$ are friction coefficients governing the exchange of momentum between the constituents. Comparing (\ref{sharppp}) to the barycentric momentum balance in dimensionless form, it turns out that the difference of $\partial_tv_i+v_i\cdot\nabla v_i$  to the mixture acceleration $\partial_tv+v\cdot\nabla v$ is negligible against the remaining terms, if the characteristic speed of diffusion is small compared to $\sqrt{p/\rho}$ which is about the speed of sound in a gas. The latter is assumed to hold, in which case (\ref{sharppp}) can be replaced by 
\begin{eqnarray*}
\rho_i(\partial_tv_i+v_i\cdot\nabla v_i)=-y_i\nabla p+y_i\lambda\nabla(\nabla\cdot v)+y_i\nabla\cdot S^\circ+\rho_i b.
\end{eqnarray*}
Applied to the solvent ($i=N$), we obtain
\begin{eqnarray}\label{crosss}
\rho^-(\partial_tv_N^-+v_N^-\cdot\nabla v_N^-)=-\nabla p_N^--RT\sum_{k=1}^{N-1}\nabla c_k^-+\lambda\nabla(\nabla\cdot v^-)+\nabla \cdot S^\circ{}^-+\rho^-b^-
\end{eqnarray}
with the standard constraint $\nabla\cdot v_N^-\equiv0$ in the incompressible limit, where  the superscript ``$-$'' indicates that a quantity refers to the liquid phase. 
For the momentum transmission,  the jump condition in (\ref{(3)}) is rewritten with $v_N$ and $j_k$, namely
\begin{eqnarray}\label{maru} 
\frac{\rho^-}{\rho_N^-}(\dot{m}_N^--j_N^-\cdot n_\Sigma)(v_N^--\frac{j_N^-}{\rho_N^-})-S^-n_\Sigma -(\dot{m}^+v^+-S^+n_\Sigma)=\nabla_\Sigma\cdot S^\Sigma\mbox{\quad on $\Sigma(t)$}.
\end{eqnarray}
In (\ref{crosss}) and (\ref{maru}), $v^-$, $S^\circ{}^-=2\eta D^\circ{}^-$ and $S^-=-p^-I+\lambda(\nabla\cdot v^-)I+2\eta D^\circ{}^-$ are to be rewritten with $v_N$ and $j_k$ instead of $v$ by means of (\ref{v_N-v}).  

In order to obtain more detailed information about the diffusive fluxes, we first compute $\nabla(\mu_i/T)$. Since we are finally interested in the isothermal case, we consider constant $T$ from here on. In the gas phase, with 
\begin{eqnarray*}  
\mu_i^+=g_i^+(T,p^R)+\frac{RT}{M_i}\ln\frac{x_i^+p^+}{p^R}=g_i^+(T,p^R)+\frac{RT}{M_i}\ln\frac{c^+_iRT}{p^R}
\end{eqnarray*}
for the assumed ideal gas mixture, we obtain the result
\begin{eqnarray*}
\nabla\frac{\mu_i^+}{T}=\frac{R}{M_i}\frac{\nabla c_i^+}{c_i^+}=\frac{R}{c_i^+}\nabla\rho_i^+,
\end{eqnarray*}
where the superscript ``$+$'' indicates that a quantity refers to the gas phase. In the liquid phase, we obtain $\nabla(\mu_i^-/T)$ for $i<N$ and $i=N$ as
\begin{eqnarray*}
\nabla\frac{\mu_i^-}{T}=\frac{R}{M_i}\frac{\nabla x_i^-}{x_i^-},\quad\nabla \frac{\mu_N^-}{T}=\frac{\nabla p^-}{T\rho_N^R}+\frac{R}{M_N}\frac{\nabla x_N^-}{x_N^-}=\frac{\nabla p_N^-}{T\rho_N^R}+\frac{R}{\rho^R_N}\sum_{k=1}^{N-1}\nabla c_k^-+\frac{R}{M_N}\frac{\nabla x_N^-}{x_N^-}.
\end{eqnarray*}
If these are inserted into the Fickean form of the diffusive mass fluxes, the (molar) mass densities in the denominator only cancel, if the dependence of the phenomenological coefficients $L_{ik}$ on $\rho_1,\ldots,\rho_N$ has a special structure. To incorporate such structural information, while keeping the derivation as rigorous as possible, we prefer to use the generalized Maxwell-Stefan equations as constitutive relations determining the diffusion fluxes. The Maxwell-Stefan equations read   
\begin{eqnarray}\label{MS}
-\sum_{k=1}^N\frac{x_kj_i^m-x_ij_k^m}{\D_{ik}}=\frac{\rho_i}{RT}\nabla\mu_i-\frac{y_i}{RT}\nabla p -\frac{\rho_i}{RT}(b_i-b)
\end{eqnarray}
with an individual body force $b_i$ for $A_i$ and the molar mass fluxes 
\begin{eqnarray*}
j_i^m:=\frac{j_i}{M_i}=c_i(v_i-v). 
\end{eqnarray*}
For a rigorous derivation of (\ref{MS}) see Bothe and Dreyer \cite{Bothe and Dreyer}. There you also find the additional contribution $\nabla \cdot S_i-y_i\nabla\cdot S$ in the right-hand side of (\ref{MS}). The latter is not included in the classical form of the Maxwell-Stefan equations as given in, e.g.,  Taylor and Krishna \cite{Taylor and Krishna} and  Bird et al.\ \cite{Bird et al.}. For simplicity, we also neglect the effect of diffusion driven by viscous stress. In (\ref{MS}), the $\D_{ik}$ are the so-called Maxwell-Stefan diffusivities, which are symmetric (cf.\ \cite{Bothe and Dreyer}). From measurements, one knows that the $\D_{ik}$ depend only weakly on the composition (often as affine functions), in contrast to the Fickean diffusivities. We assume the $\D_{ik}$ to be constant with $\D_{ik}=\D_{ki}>0$. Note that the Maxwell-Stefan equations sum up to zero, and hence the N equations are not independent. Concerning the inversion of this equation system, see Bothe \cite{Bothe}. 
 
From here on, we assume equal body forces $b_k\equiv b$ for all components. Insertion of the chemical potential gradients yields for the gas phase
\begin{eqnarray*} 
-\sum_{k\neq i}\frac{x_k^+j_i^m{}^+-x_i^+j_k^m{}^+}{\D_{ik}^+}=\nabla c_i^+-\frac{y_i^+}{RT}\nabla p^+ =\nabla c_i^+-y_i^+\nabla c^+.  
\end{eqnarray*} 
In the liquid phase, we obtain for $i<N$ 
\begin{eqnarray*} 
-\sum_{k\neq i}\frac{x_k^-j_i^m{}^--x_i^-j_k^m{}^-}{\D_{ik}^-}=c^-\nabla x_i^-- \frac{y_i^-}{RT}\nabla p^-.  
\end{eqnarray*} 
 For $i=N$, we obtain 
 \begin{eqnarray*} 
-\sum_{k\neq N}\frac{x_k^-j_N^m{}^--x_N^-j_k^m{}^-}{\D_{Nk}^-}=\frac{\nabla p^-}{RT}+c^-\nabla x_N^-- \frac{y_N^-}{RT}\nabla p^-.  
\end{eqnarray*} 
\indent We simplify the jump conditions of the chemical potential. Neglecting the viscous and the kinetic effect in (B7), we assume
\begin{eqnarray*} 
 [\![\mu_i]\!]=0.
\end{eqnarray*}
See Bothe and Fleckenstein \cite{Bothe and Fleckenstein} for an assessment of this approximation. 
For $i<N$, we have
\begin{eqnarray*}
\mu_i^+(T,\rho^+,x'{}^+)&=&g_i^+(T,p^R)+\frac{RT}{M_i}\ln\frac{p^+x_i^+}{p^R}=g_i^+(T,p^R)+\frac{RT}{M_i}\ln \frac{p_i^+}{p^R},\\
\mu_i^-(T,\rho^-,x'{}^-)&=&g_i^-(T,p^R)+\frac{RT}{M_i}\ln x_i^-. 
\end{eqnarray*}
For given $T$, choose $p^R=p_i^R(T)$ so that $g_i^+(T,p_i^R(T))=g_i^-(T,p_i^R(T))$ holds for each $i$ and for a planar interface. Then, neglecting curvature effects via the pressure jump,  we obtain 
\begin{eqnarray*}
\mu_i^+(T,\rho^+,x'{}^+)=\mu^-_i(T,\rho^-,x'{}^-)\Leftrightarrow \ln x_i^-=\ln \frac{p^+_i}{p^R_i(T)}=\ln \frac{\rho^+_i RT}{M_ip_i^R(T)}\Leftrightarrow x_i^-p^R_i(T)=c_i^+RT.
\end{eqnarray*}
This is a version of Henry's law. Thus we obtain the following PDE system for incompressible solvent and compressible gas phase. 
\medskip \medskip

\noindent {\bf  Non-dilute solution with incompressible solvent:}

\noindent Gas phase:
\begin{eqnarray*}
\left\{
\begin{array}{lll}
&\dis\partial_t c_i+ \nabla\cdot(c_iv+j_i^m)=0,\qquad i=1,\ldots,N, \medskip\\
&\dis -\sum_{k\neq i}\frac{x_kj_i^m-x_ij_k^m}{\D_{ik}}=\nabla c_i-y_i\nabla c,\qquad i=1,\ldots,N, \medskip\\
&\dis \rho(\partial_tv+v\cdot\nabla v)+\nabla p =\lambda \nabla (\nabla \cdot v)+\eta\Delta v + \rho b,\medskip\\
&\dis p=cRT=RT\sum_{k=1}^Nc_k,\,\,\,\rho=\sum_{k=1}^NM_kc_k.
\end{array}
\right.
\end{eqnarray*}

\noindent Liquid phase:
\begin{eqnarray*}
\left\{
\begin{array}{lll}
&\dis \partial_t c_i+ \nabla\cdot(c_iv_N+J_i^m)=0,\qquad i=1,\ldots,N-1, \medskip\\
&\dis -\sum_{k\neq i}\frac{x_kj_i^m-x_ij_k^m}{\D_{ik}}=c\nabla x_i-\frac{y_i}{RT}\nabla p,\qquad i=1,\ldots,N-1, \medskip\\
&\dis -\sum_{k\neq N}\frac{x_kj_N^m-x_Nj_k^m}{\D_{Nk}}=c\nabla x_N+\frac{1-y_N}{RT}\nabla p, \medskip\\
&\dis J_i^m=j_i^m+\frac{\rho_i}{\rho_N}\sum_{k=1}^{N-1}j^m_k,\,\,\,j_i^m=\frac{j_i}{M_i},\,\,\,p=p_N+\sum_{k=1}^{N-1}c_kRT,\medskip\\
&\dis \rho(\partial_tv_N+v_N\cdot\nabla v_N)+\nabla p_N = \lambda\nabla(\nabla \cdot v) +\nabla \cdot S^\circ + \rho b -RT\sum_{k=1}^{N-1}\nabla c_k, \medskip \\
&\dis \nabla\cdot v_N\equiv0,\,\,\,\rho_N\equiv\rho_N^R,
\end{array}
\right.
\end{eqnarray*}
where $v$ and $S^\circ=2\eta D^\circ=\eta \big(\nabla v+(\nabla v)^{\sf T}-\frac{1}{3}(\nabla\cdot v)I\big)$ are to be rewritten with $v_N$ and $j_k$ through (\ref{v_N-v}). 

\noindent Interface:
\begin{eqnarray*}
\left\{
\begin{array}{lll}
&\dis(j_i^m{}^++c_i^+(v^+-v^\Sigma))\cdot n_\Sigma=(j_i^m{}^-+c_i^-(v_N^--v^\Sigma))\cdot n_\Sigma, \qquad i=1,\ldots,N-1, \medskip\\
&\dis (j^+_N\cdot n_\Sigma+y_N^+\rho^+(v^+-v^\Sigma))\cdot n_\Sigma=\rho^-_N(v^-_N-v^\Sigma)\cdot n_\Sigma \medskip\\ 
& \mbox{(or  $v^\Sigma\cdot n=v^-_N\cdot n_\Sigma$ in case of negligible evaporation of $A_N$)}, \medskip\\
&\dis x_i^-p_i^R(T)=c_i^+RT,\qquad i=1,\ldots,N-1,\medskip\\
&\dis \frac{\rho^-}{\rho_N^-}(\dot{m}_N^--j_N^-\cdot n_\Sigma)(v_N^--\frac{j_N^-}{\rho_N^-})-S^-n_\Sigma -(\dot{m}^+v^+-S^+n_\Sigma)=\gamma^\Sigma\kappa_\Sigma n_\Sigma+\nabla_\Sigma\gamma^\Sigma,
\end{array}
\right.
\end{eqnarray*}
where  $\kappa_\Sigma=\nabla_\Sigma\cdot(-n_\Sigma)$ is the curvature, $S^\pm=-p^\pm I+\lambda(\nabla\cdot v)^\pm I+2\eta D^\circ{}^\pm$ with $D^\circ=\frac{1}{2}\big(\nabla v+(\nabla v)^{\sf T}-\frac{1}{3}(\nabla\cdot v)I\big)$ and $S^-$ is to be rewritten with $v_N$ and $j_k$ through (\ref{v_N-v}).
\setcounter{section}{5}
\setcounter{equation}{0}
\section{Dilute Solution with Incompressible Solvent}

\noindent We ignore bulk viscosities in both phases and assume $\dot{m}_N=0$. Note that in the dilute case ($x_i\ll 1$ for $i<N$), we  have $J_i^m\approx j_i^m$, $\rho\approx\rho_N$ and $S\approx S_N$ in the liquid phase. We obtain a simple Fick's law for $i<N$, namely  
\begin{eqnarray*}
j_i^m{}^-= -\D_{Ni}^-\nabla c_i^-. 
\end{eqnarray*} 
We may approximate $c^-\approx c_N^-$. Then Henry's law becomes  
\begin{eqnarray*} 
\frac{c^-_i}{c_i^+}=\frac{c_N^-RT}{p^R_i(T)}=:H_i.
\end{eqnarray*}
Thus we obtain the following PDE system for a dilute solution with incompressible solvent and compressible gas phase. 
\medskip \medskip

\noindent {\bf  Dilute solution with incompressible solvent:}

\noindent Gas phase: 
\begin{eqnarray*} 
\left\{
\begin{array}{lll}
&\dis \partial_t c_i+ \nabla\cdot(c_iv+j_i^m)=0,\qquad i=1,\ldots,N,\medskip\\
&\dis -\sum_{k\neq i}\frac{x_kj_i^m-x_ij_k^m}{\D_{ik}}=\nabla c_i-y_i\nabla c,,\qquad i=1,\ldots,N, \medskip\\
&\dis \rho(\partial_tv+v\cdot\nabla v)+RT\nabla c =\eta\Delta v + \rho b,\medskip\\
&\dis p=cRT=RT\sum_{k=1}^Nc_k,\quad \rho=\sum_{k=1}^NM_kc_k.
\end{array}
\right.
\end{eqnarray*}

\noindent Liquid phase:
\begin{eqnarray*}
\left\{
\begin{array}{lll}
& \partial_t c_i+ \nabla\cdot(c_iv_N+j_i^m)=0,\qquad i=1,\ldots,N-1,\medskip \\
&j_i^m=-\D_{iN}\nabla c_i,\qquad i=1,\ldots,N-1,\medskip \\
&\dis \rho_N(\partial_tv_N+v_N\cdot\nabla v_N)+\nabla p_N =\eta_N\Delta v_N + \rho_N b-RT\sum_{k=1}^{N-1}\nabla c_k,\medskip \\
&\nabla\cdot v_N\equiv0,\,\,\,\rho_N\equiv\rho_N^R.
\end{array}
\right.
\end{eqnarray*}

\noindent Interface:
\begin{eqnarray*}
\left\{
\begin{array}{lll}
&(j_i^m{}^++c_i^+(v^+-v^\Sigma))\cdot n_\Sigma=j_i^m{}^-\cdot n_\Sigma,\qquad i=1,\ldots,N-1,\medskip\\
&\dis v^\Sigma\cdot n=v^-_N\cdot n_\Sigma,\medskip\\
&\dis \frac{c_i^-}{c_i^+}=H_i,\medskip\\
&\dis (-j_N^-\cdot n_\Sigma)(v_N^--\frac{j_N^-}{\rho_N^-})-S_N^-n_\Sigma -(\dot{m}^+v^+-S^+n_\Sigma)=\gamma^\Sigma\kappa_\Sigma n_\Sigma+\nabla_\Sigma\gamma^\Sigma,
\end{array}
\right.
\end{eqnarray*}
where $S^+=-p^+I+2\eta D^\circ{}^+$ and $S^-_N=-p_N^-I+2\eta_N D^\circ{}^-=-p_N^-I+\eta^-_N(\nabla v_N+(\nabla v_N)^{\sf T})^-$.
\setcounter{section}{6}
\setcounter{equation}{0}
\section{Boundary Conditions}
 The mathematical model is to be complemented by appropriate boundary conditions at the fixed walls, called $\partial \Omega$, and at the free upper surface $\Gamma(t)$. Since the derivation of physically  sound boundary conditions is a topic on its own (cf., e.g.\  Bothe, K\"ohne and Pr\"uss \cite{Bothe Kohne and Pruss}), we rest content with the simplest reasonable choice. 

\noindent{\bf (i) Boundary conditions at fixed walls}

\noindent The fixed walls are impermeable. Hence
\begin{eqnarray*}
\mbox{ $v\cdot n_w=0$ \,\,and\,\, $j_i\cdot n_w=0$ \,\,at  $\partial \Omega$,}
\end{eqnarray*}
where $n_w$ is the unit outer normal to the walls. 
\noindent This also implies 
\begin{eqnarray*}
\mbox{$v_N\cdot n_w=0$\,\, at  $\partial \Omega$.}
\end{eqnarray*}
In order to allow for a movable upper surface, the tangential velocities $v_\parallel$ and $v_N{}_\parallel$ shall not be assumed to vanish. Instead, we assume a Navier slip condition of the form 
\begin{eqnarray*}
\mbox{$v_\parallel+a(Sn_w)_\parallel=0$,  \,\,$v_N{}_\parallel+a_N(S_Nn_w)_\parallel=0$ \,\, at  $\partial\Omega$  with $a,a_N\ge0$.} 
\end{eqnarray*}
 In the non-isothermal case, we add a Robin-condition for the temperature, i.e.
\begin{eqnarray*}
\mbox{$T+\beta\nabla T\cdot n_w=T_{ext}$ \,\, at  $\partial \Omega$}
\end{eqnarray*}
with $\beta\ge0$.

\noindent {\bf (ii) Boundary conditions at the free upper surface}

\noindent The Robin condition for the temperature can also be applied at the free surface. The other conditions are 
\begin{eqnarray}\label{BC5}
\mbox{$v\cdot n_\Gamma=V_\Gamma$\, and\, $(p_{ext}-p)n_\Gamma+S^{irr}n_\Gamma=\gamma^\Gamma \kappa_\Gamma n_\Gamma+\nabla_\Gamma\gamma^\Gamma$ \,\, on  $\Gamma(t)$}
\end{eqnarray}
with the outer unit normal $n_\Gamma$ on $\Gamma(t)$ and  the curvature $\kappa_\Gamma=\nabla_\Gamma\cdot(-n_\Gamma)$. 
Let us note that, in the dilute solution limit and for a constant surface tension $\gamma^\Gamma$, the condition (\ref{BC5}) becomes 
\begin{eqnarray*}
\mbox{$v_N\cdot n_\Gamma=V_\Gamma$,\,\,\,  $p_{ext}-p_N+n_\Gamma\cdot S^{irr}_Nn_\Gamma=\gamma_\Gamma \kappa_\Gamma$ \,\,and\,\, $n_\Gamma\times S_N^{irr}n_\Gamma=0$ \,\, on  $\Gamma(t)$.}
\end{eqnarray*}
We assume the mixture composition to be given at $\Gamma(t)$ due to local chemical equilibrium with a large and well-mixed external gas phase. Hence 
\begin{eqnarray*}
\mbox{$x_i^-=x_i^\Gamma$\,\,\, on $\Gamma(t)$ for $i<N$ with $x_i^\Gamma\ge0$,} 
\end{eqnarray*}
where we assume $\sum_{i=1}^{N-1}x_i^\Gamma<1$.

\noindent {\bf (iii) Condition at the contact line}

\noindent The free surface $\Gamma(t)$ touches the fixed wall in a set of points which forms the so-called contact line $C$. The modeling of dynamic contact lines is, again, a topic on its own and we refer to Shikhmurzaev \cite{Shikhmurzaev} and the reference therein for detailed information. Here, in order to close the system in the simplest possible manner, we assume a fixed contact angle of $\pi/2$, i.e.
\begin{eqnarray*}
\mbox{$n_\Gamma\perp n_w$ \,\, on $C$.}
 \end{eqnarray*}

\appendix
\def\thesection{Appendix}
\section{}
The derivation of the balance equations is based on standard two-phase transport and divergence theorems: Let $V$ denote an arbitrary fixed control volume with outer normal $n$. Then   
\begin{eqnarray*}
\frac{d}{dt}\int _V\phi dx = \int_{V\setminus \Sigma} \partial_t \phi dx -\int_{\Sigma_V}[\![\phi]\!]v^\Sigma\cdot n_\Sigma do
\end{eqnarray*}
with $\Sigma_V:=\Sigma(t)\cap V$, the surface velocity (including tangential part) $v^\Sigma$ and the surface unit normal $n_\Sigma$.  Here $[\![\phi]\!]:=\lim_{h\to0+}(\phi(x+hn_\Sigma)-\phi(x-hn_\Sigma) )$ defined for $x\in\Sigma$. We also have 
\begin{eqnarray*}
\int_{\partial V}f \cdot n do=\int_{V\setminus \Sigma} \nabla \cdot f dx+\int_{\Sigma_V}[\![ f\cdot n_{\Sigma}]\!]do. 
\end{eqnarray*}  
Since the internal energy and the entropy have surface contributions, we also need the surface transport theorem for $\phi^\Sigma$ defined on $\Sigma$. It states that    
\begin{eqnarray*}
\frac{d}{dt}\int_{\Sigma_V}\phi^\Sigma do =\int_{\Sigma_V}\left( \frac{D^\Sigma\phi^\Sigma}{Dt}+\phi^\Sigma \nabla_\Sigma\cdot v^\Sigma \right)do -\int_{\partial \Sigma_V} \phi^\Sigma v^\Sigma\cdot \nu ds,  
\end{eqnarray*} 
which -- in this simple form -- holds for all fixed $V$ such that its outer normal $n$ satisfies $n\perp n_\Sigma$ on $\Sigma_V$, and hence $n=\nu$, where $\nu$ is tangential to $\Sigma$ and normal to the bounding curve $\partial \Sigma_V$. We always choose such control volumes in the integral balances above. For more details and mathematical proofs see, e.g., Slattery \cite{Slattery}, Romano and Marasco \cite{Romano and Marasco} or the appendix in Bothe, Pr\"uss and Simonett \cite{Bothe-Pruss-Simonett}.


\end{document}